\documentclass[intlimits,twoside,a4paper]{article}

\usepackage{amsmath,amssymb}
\usepackage{graphicx}
\usepackage{grffile}
\usepackage{amsfonts}
\usepackage{hhline}
\usepackage{booktabs}
\usepackage{array}

\usepackage[T2A]{fontenc}
\usepackage[cp1251]{inputenc}

\newcommand{\RN}[1]{%
  \textup{\uppercase\expandafter{\romannumeral#1}}%
}
\newcommand{\overbar}[1]{\mkern 1.5mu\overline{\mkern-1.5mu#1\mkern-1.5mu}\mkern 1.5mu}

\usepackage[eqsecnum]{cmpj2}

\issue{2016}{19}{4}{43601}
\doinumber{10.5488/CMP.19.43601}

\title[Structural, mechanical and thermal properties of Cu$_2$CdSnSe$_4$ and Cu$_2$HgSnSe$_4$ adamantine materials]%
{Computational insight on the structural, mechanical and thermal properties of Cu$_2$CdSnSe$_4$ and Cu$_2$HgSnSe$_4$ adamantine materials}
\author[S. Bensalem, M. Chegaar, A. Bouhemadou]{S.~Bensalem\refaddr{label1}\thanks{Corresponding author, E-mail: bensalemse@gmail.com.}\,,
M.~Chegaar\refaddr{labe12,labe22},
A.~Bouhemadou\refaddr{labe13}}
\addresses{
\addr{label1} Centre de D\' eveloppemen des Energies Renouvelables, CDER, BP 62 Route de l'Observatoire Bouzar\' eah, \\16340 Algiers, Algeria
\addr{labe12} D\' epartement de Physique, Facult\' e des Sciences, Universit\' e S\' etif 1, 19000 S\' etif, Algeria
\addr{labe22} Laboratoire d'Opto\' electronique et Composants, Universit\' e S\' etif 1, 19000 S\' etif, Algeria
\addr{labe13} Laboratory for Developing New Materials and their Characterization, University of Setif 1, Setif 19000, Algeria}
\authorcopyright{S. Bensalem, M. Chegaar, A. Bouhemadou, 2016}
\date{Received March 24, 2016, in final form June 27, 2016}

\begin{document}

\maketitle

\begin{abstract}
Through first-principles calculation based on the density functional theory (DFT) within the pseudo potential-plane wave (PP-PW) approach, we studied the structural, mechanical and thermal properties of Cu$_2$CdSnSe$_4$ and Cu$_2$HgSnSe$_4$ adamantine materials. The calculated lattice parameters are in good agreement with experimental and theoretical reported data. The elastic constants are calculated for both compounds using the static finite strain scheme. The hydrostatic pressure action on the elastic constants predicts that both materials are mechanically stable up to 10~GPa. The polycrystalline mechanical parameters, i.e., the anisotropy factor ($A$), bulk modulus ($B$), shear modulus ($G$), Young's modulus ($E$), Lame's coefficient ($\lambda$) and Poisson's ratio ($\nu$) have been estimated from the calculated single crystal elastic constants. The analysis of $B/G$ ratio shows that the two studied compounds behave as ductile. Based on the calculated mechanical parameters, the Debye temperature and the thermal conductivity have been probed. In the framework of the quasi-harmonic approximation, the temperature dependence of the lattice heat capacity of both crystals has been investigated.
\pacs 63.20.dk, 62.20.D-, 62.20.fk, 65.40.Ba
\end{abstract}

\section{Introduction}

Nowadays, thermoelectric effect is gaining significant consideration due to its relevance to \linebreak green energy conversion and sustainable development \cite{Zhang201592}. The copper-based adamantine compounds \linebreak Cu$_2$\RN{2}SnSe$_4$ (\RN{2}= Zn, Cd, Hg) represent a promising family of p-type semiconductors for excellent thermoelectric power generation \cite{Navrtil2014, Fan2011, Liu2014, Li2013}. The conversion efficiency of a thermoelectric material is related to the dimensionless figure of merit, $ZT$, which is related to the Seebeck coefficient ($S$), the electrical conductivity ($\sigma$), the thermal conductivity ($\kappa$) and the absolute temperature ($T$) by \cite{Zhang201592}:
\begin{eqnarray}
  ZT=\frac{S^2\sigma}{\kappa}T.
\end{eqnarray}

The achievement of high values of ($ZT$) is a primordial task in thermoelectric research activities. Fan and co-workers reported peak values of the dimensionless figure of merit at 450\textcelsius; $ZT$=0.44 for Cu$_2$ZnSnSe$_4$ \cite{Fan2012} and 0.65 for Cu$_2$CdSnSe$_4$ \cite{Fan2011}. Navratil et al. \cite{Navrtil2014} found that Cu$_2$HgSnSe$_4$ sample exhibits a reasonable value of $ZT$=0.3 at 600~K. In fact, these compounds are attractive candidates for advanced thermoelectric devices.

In the literature, most theoretical contributions on Cu$_2$ZnSnSe$_4$ (CZTSe), Cu$_2$CdSnSe$_4$ (CCTSe) and Cu$_2$HgSnSe$_4$ (CHTSe) materials mainly consider their electronic and optical properties. Botti et al. \cite{Botti2011} investigated the band structure of CZTSe through many-body methods, their calculations are based on a restricted self-consistent GW scheme, and the estimated band gap is 0.86~eV. Chen et al. \cite{Chen2009} calculated the electronic band structure of CZTSe with VASP code, the authors have performed hybrid functional calculations (HSE06) yielding the value of 0.82~eV as a theoretical gap. Persson \cite{Persson2010} studied the electronic and optical properties of CZTSe by means of WIEN2K code, in his contribution the author has used an optimized GGA+U as exchange correlation potential, which yields a gap value of 0.89~eV for CZTSe. Nakamura et al. \cite{Nakamura2011} reported the phase stability and electronic structures of CCTSe and CHTSe compounds via first-principles calculations using CASTEP code. In their study, the screened-exchange LDA functional has been used and the estimated band gaps are 0.52~eV and 0.07~eV for CCTSe and CHTSe, respectively. Li et al. \cite{LiD2012} presented an ab-initio contribution of structural, electronic and optical properties of CHTSe using CASTEP software program, and they have found that this crystal is a direct gap semiconductor at the $\Gamma$ point. Nevertheless, some points bearing on mechanical and thermal properties do not appear to have been previously reported or are still not clear for these compounds.

In a previous work \cite{Bensalem2014}, we performed a theoretical study of structural, mechanical and thermodynamic properties of Cu$_2$ZnSnSe$_4$. Hence, this study deals with both remaining materials in order to fully take advantage of this interesting family of compounds for eventual technological applications, for example as thermoelectric materials. For this purpose, we bring out numerical results on the structural parameters, mechanical characteristics and thermal properties of Cu$_2$CdSnSe$_4$ and Cu$_2$HgSnSe$_4$ adamantine compounds.

To exhibit our work, the layout of this paper is organized as follows: in section~\ref{2}, we present the computational details of the employed calculation techniques. In section~\ref{3}, the discussion of the obtained results will be presented; including the structural parameters, the elastic constants with some related mechanical properties and the thermal properties, namely, the Debye temperature, the thermal conductivity and the heat capacity. We conclude by summarizing the obtained main results in section~\ref{4}.

\section{Computational details}
\label{2}
All our calculations were performed using CASTEP (Cambridge Sequential Total Energy Package) \cite{CASTEP} version 7.0. In this code, the pseudo potential-plane wave (PP-PW) approach \cite{CASTEP, RMP-Payne} within the density functional theory (DFT) \cite{DFT-HK, DFT-KS} formalism is implanted. Here, we used a PW basis set defined by an energy cut-off 400~eV, such a value ensuring the convergence of the total energy and yielding lattice parameters that have better agreement with experimental data. Interactions of electrons with ion cores are represented by the Vanderbilt-type ultrasoft PP \cite{Vanderbilt1990}. The states Cu~(3d$^{10}$~4s$^{1}$), Cd~(4d$^{10}$~5s$^{2}$), Hg~(5d$^{10}$~6s$^{2}$), Sn~(5s$^{2}$~5p$^{2}$) and Se~(4s$^{2}$~4p$^{4}$) were treated as valence states. The electronic exchange correlation interactions are treated within the Wu-Cohen generalized gradient approximation (GGA-WC) \cite{Wu2006}. The Brillouin zone sampling was carried out using 75~k-points in the irreducible part, which correspond to 6\texttimes 6\texttimes 7 set of Monkhorst-Pack points \cite{Monkhorst1976} for both studied materials.

The structural parameters were determined using the Broyden-Fletcher-Goldfarb-Shenno (BFGS) mi\-nimization technique \cite{BFGS}. The system reaches the ground state via self-consistent calculation when: the total energy is stable within 5\texttimes $10^{-6}$~eV/atom, the displacement of atoms during the geometry optimization is less than 5\texttimes $10^{-4}$~\AA, the maximum ionic Hellmann-Feynman force is less than $10^{-2}$~eV/{\AA} and the maximum stress is within 2\texttimes $10^{-2}$~GPa.

The elastic constants were determined by applying a set of given homogeneous deformations with a finite value and by calculating the resulting stress with respect to optimizing the internal atomic freedoms \cite{Milman2001}. The criteria for convergences of optimization on atomic internal freedoms were chosen as follows: the difference of total energy within $10^{-6}$~eV/atom, ionic Hellmann-Feynman force within 2\texttimes $10^{-3}$~eV/{\AA} and maximum ionic displacement within $10^{-4}$~\AA. The maximum strain amplitude was set to be 3\texttimes $10^{-3}$ in the present fundamental inquiry.

Phonon calculation was used to evaluate the temperature dependence of the lattice heat capacity for both considered CCTSe and CHTSe crystals in the framework of quasi-harmonic approximation. CASTEP code allows us to perform a phonon calculation using finite difference method. In the finite difference scheme, a super-cell defined by cutoff radius in {\AA} specifies the real space cutoff radius for dynamical matrix calculations. This introduced value of cutoff radius will be used to construct an appropriate super-cell. In the present theoretical study, we choose a cutoff radius of 5~{\AA} yielding a super-cell with a volume which is 8 times the original cell.

\section{Results and discussion}
\label{3}
\subsection{Structural parameters}

\begin{figure}[!h]
\centerline{\includegraphics[width=0.8\textwidth]{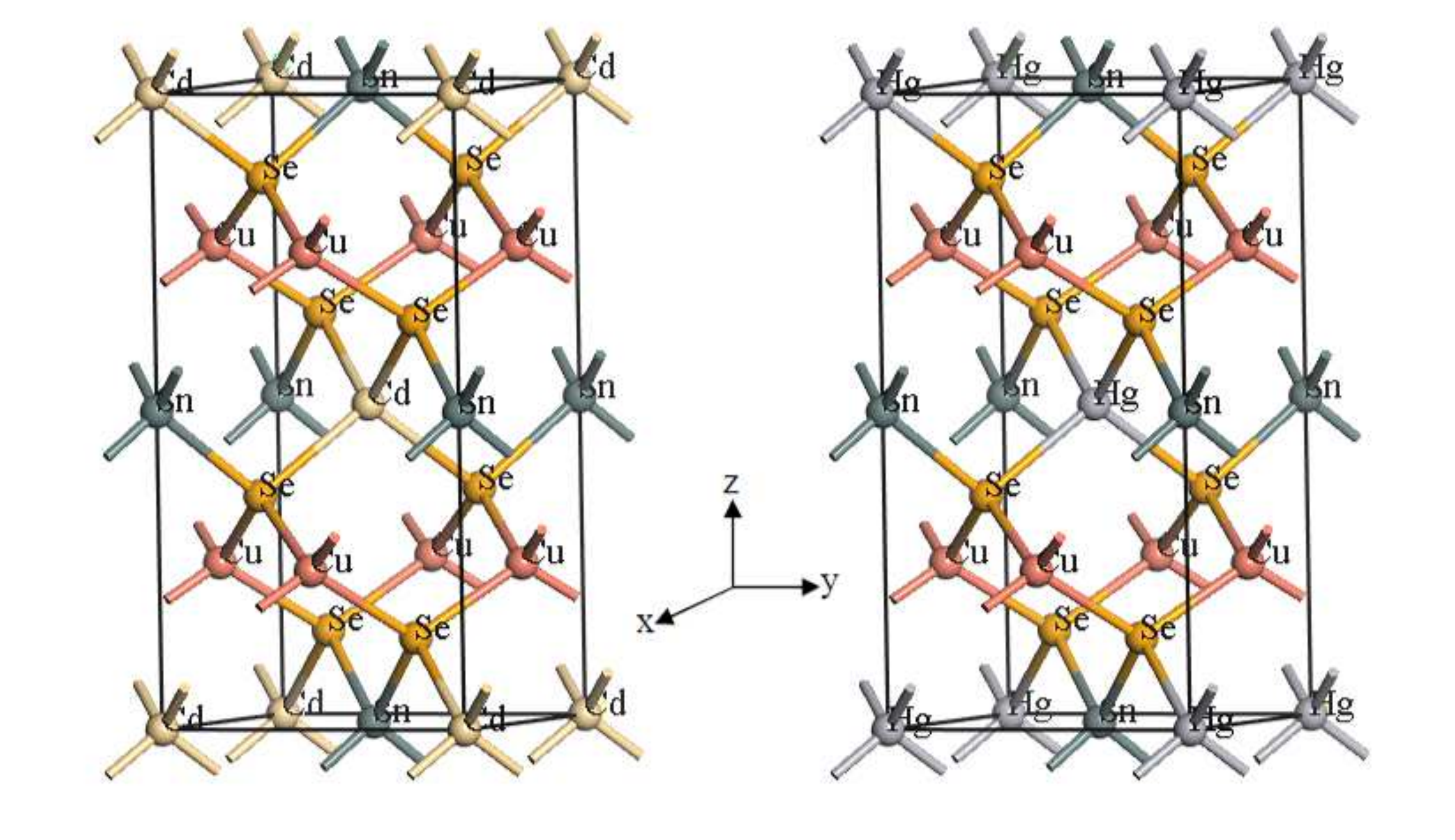}}
\caption{(Color online) Ball and stick representation of the conventional cell structure of Cu$_2$CdSnSe$_4$ (left) and Cu$_2$HgSnSe$_4$ (right) crystals in the stannite phase.}
\label{fig-1}
\end{figure}

The most energetically favorable structure for CCTSe and probably CHTSe is that of stannite (ST)~\cite{Li2013}. The stannite-type belongs to the adamantine family, which is derived from diamond structure \cite{Pamplin1980}. Figure~\ref{fig-1} displays the crystalline structure of both studied materials considering their conventional cells. The stannite Cu$_2$\RN{2}SnSe$_4$ (\RN{2}= Cd, Hg) with space group  (I\={4}2m; no. 121) has four equivalent Cu atoms on Wyckoff 4d position (0, 1/2, 1/4), two \RN{2} atoms on 2a site (0, 0, 0), two Sn atoms on 2b (0, 0, 1/2) and eight Se atoms on the 8i position defined by $(x/a, x/a, z/c)$ internal coordinates.

Considering their primitive cells, the total energies provided by the BFGS scheme are: $E_{\text{tot}}$(CCTSe)$=-5369.1103$~eV and $E_{\text{tot}}$(CHTSe)$=-5215.5058$~eV. These data may be used to investigate the ground state of these materials comparing them with eventual other phases.

Table~\ref{tab:1} illustrates the calculated lattice parameters at ambient pressure, a good agreement with experimental and theoretical reported data is obvious; the deviations between the experimental and the calculated conventional cell parameters $a$ and $c$ are estimated by less than 0.75\%. These deviations show that the present calculations are highly reliable. Consequently, the usage of the optimized structure is entirely reasonable to perfectly perform the subsequent calculations.
\begin{table}[!t]
\centering
\caption{Calculated lattice parameters (in \AA), $c/a$ ratio and the conventional-cell volume $V_0$ (in \AA$^3$) compared with experimental and theoretical available data.}
\label{tab:1}\vspace{2ex}
\begin{tabular}{|l|l|l|l|}
\hline\hline
&        Present work & Experiment \cite{Olekseyuk2002} & Theoretical \cite{Nakamura2011}\\
\hline\hline
\multicolumn{4}{|c|}{CCTSe}             \\ \hline
$a$      & 5.792  & 5.834      & 5.879      \\ \hline
$c$      & 11.384 & 11.404     & 11.514     \\ \hline
$c/a$    & 1.965  & 1.955      & $-$          \\ \hline
$V_0$    & 381.90 & 387.20     & $-$      \\
\hline
\multicolumn{4}{|c|}{CHTSe}       \\ \hline
$a$      & 5.810  & 5.829      & 5.888  \\\hline
$c$      & 11.399 & 11.418     & 11.558 \\\hline
$c/a$    & 1.962  & 1.959      & $-$\\\hline
$V_0$    & 384.79 & 388.62     & $-$ \\
\hline\hline
\end{tabular}
\end{table}

\subsection{Mechanical characteristics}

The calculated elastic constants ($C_{ij}$) are listed in table~\ref{tab:2}. All elastic constants for the herein studied compounds are positive and satisfy the criteria for mechanically stable crystals, which are given in the case of tetragonal symmetry by \cite{Wallace1972}:
\begin{eqnarray}
\left\{
\begin{array}{l}
C_{11}>0,\qquad C_{44}>0,\qquad C_{66}>0,\\
C_{11}-C_{12}>0,\\
C_{33}(C_{11}+C_{12})-2C_{13}^{2}>0.
\end{array}
\right.
\end{eqnarray}
\begin{table}[!t]
\centering
\caption{Calculated elastic constants $C_{ij}$ (in GPa) for the ST-type of CCTSe and CHTSe adamantine compounds.}
\label{tab:2}\vspace{2ex}
\begin{tabular}{|l|l|l|l|l|l|l|l|}
\hline\hline
&       $C_{11}$  & $C_{33}$ & $C_{44}$ & $C_{66}$ & $C_{12}$ & $C_{13}$       \\
\hline\hline
CCTSe  & 82.398    & 79.017   & 28.599   & 31.945   & 56.593   & 55.523         \\ \hline
CHTSe  & 80.983    & 79.617   & 30.762   & 33.358   & 54.783   & 54.364         \\
\hline\hline
\end{tabular}
\end{table}

To probe the mechanical stability of a crystal at high pressure, the study of the elastic constants evolution under pressure gradient is required. The generalized Born stability criteria for the tetragonal systems under an external hydrostatic pressure, $P$, is given by \cite{Manjon2014, Wallace1972}:
\begin{eqnarray}
\left\{
\begin{array}{l}
C_{11}-P>0,\qquad C_{44}-P>0,\qquad C_{66}-P>0,\\
C_{11}-C_{12}-2P>0,\\
(C_{33}-P)(C_{11}+C_{12})-2(C_{13}+P)^{2}>0.
\end{array}
\right.
\end{eqnarray}

\begin{figure}[!t]
\vspace{-2ex}%
\centerline{\includegraphics[width=0.7\textwidth]{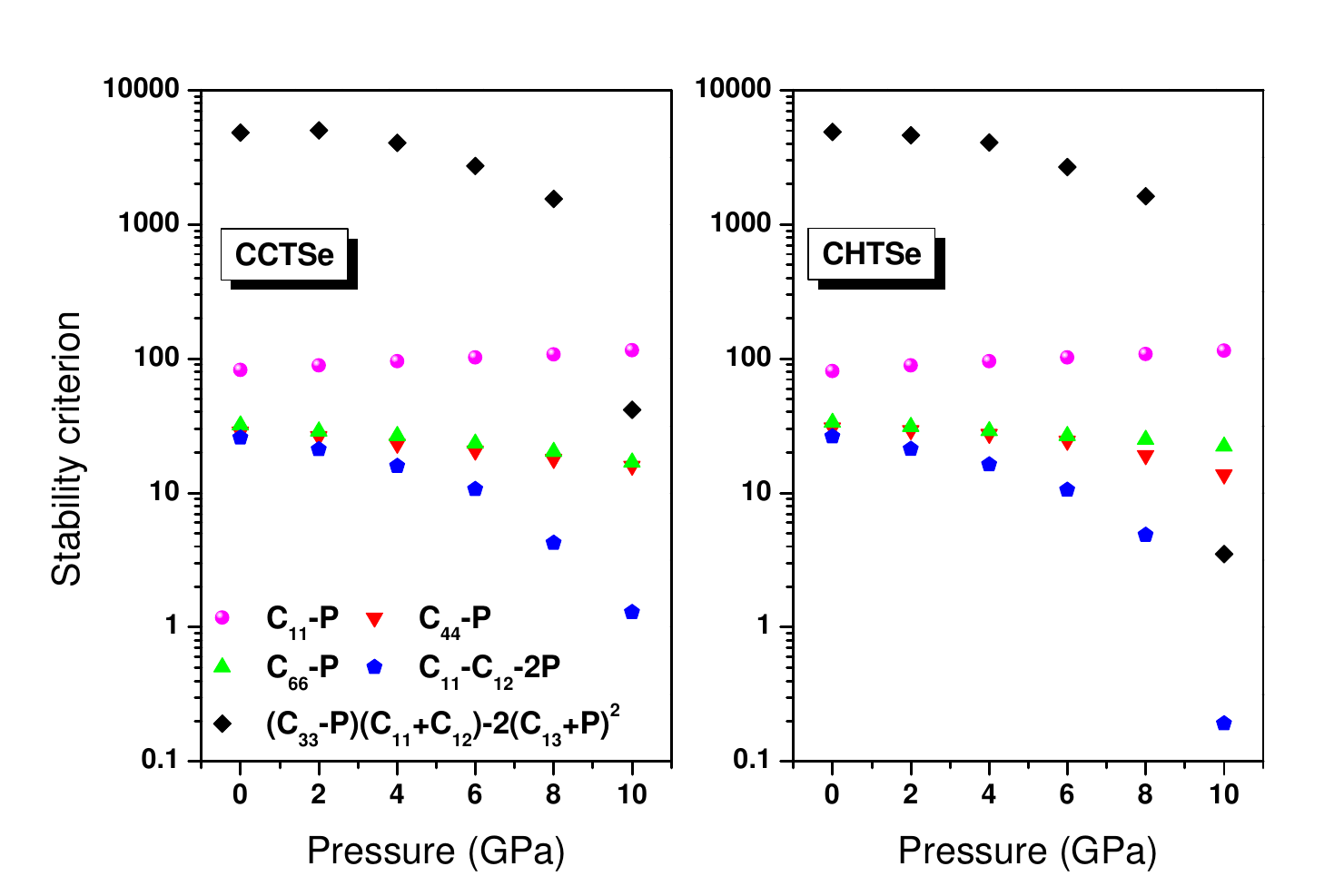}}
\caption{(Color online) The generalized criteria of stability in the range 0--10~GPa. Taking into account the large range of the orders of magnitude, and to show that the criteria are strictly positives, the logarithmic scale is adopted here.}
\label{fig-2}
\end{figure}

Figure~\ref{fig-2} displays the evolution of these generalized criteria of stability in the range of the hydrostatic pressure that we have opted in our study. Figure~\ref{fig-2} shows that both studied compounds satisfy the generalized criteria in the pressure range 0--10~GPa. Accordingly, CCTSe and CHTSe are mechanically stable up to 10~GPa.

\begin{figure}[!t]
\vspace{-2ex}%
\centerline{\includegraphics[width=0.7\textwidth]{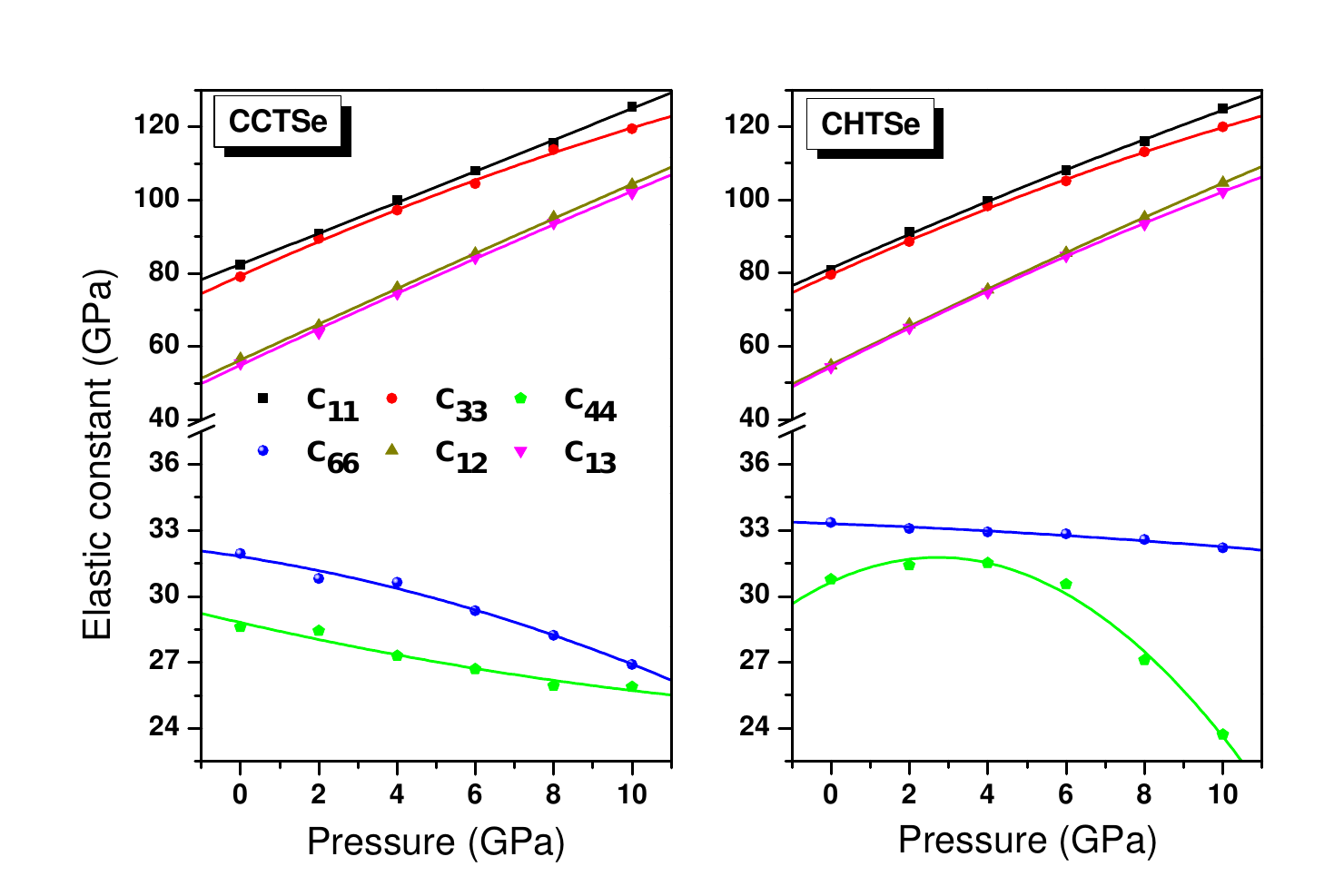}}
\caption{(Color online) The predicted pressure effect on the elastic constants of CCTSe and CHTSe. The symbols are the calculated values. However, the solid lines represent the second-order polynomial fit based on equation~\ref{Cij-P} with the fitting parameters listed in table~\ref{tab:3}.}
\label{fig-3}
\end{figure}

Figure~\ref{fig-3} illustrates the evolution of the elastic constants with respect to the variation of pressure. Practically all the elastic constants exhibit a monotonous behavior, except $C_{44}$ in the case of Hg-based compound. It increases in the onset of the pressure range and decreases above 3~GPa. This elastic constant may vanish at pressure value slightly greater than 10~GPa, suggesting an elastic instability in the case of Hg-based compound compared with the Cd-based one at the upper neighborhood of 10~GPa, similar analyses in the same context about the behavior of the elastic constants being available in \cite{Karki2001, Stixrude1998}.

Table~\ref{tab:3} lists the values of pressure coefficients provided by fitting the calculated data points to second-order polynomials:
\begin{equation}
\label{Cij-P}
C_{ij}(P)=C_{ij0}+\alpha P+\beta P^2,
\end{equation}
here, $\alpha$ and $\beta$ are the pressure coefficients; $C_{ij0}$ designates the value of the elastic constant at zero pressure.
\begin{table}[!t]
\centering
\caption{Pressure coefficients provided by the performed fit based on equation~(\ref{Cij-P}).}
\label{tab:3}\vspace{2ex}
\begin{tabular}{|l|l|l|l|l|}
\hline\hline
& \multicolumn{2}{c|}{CCTSe} & \multicolumn{2}{c|}{CHTSe}\\
\hline
&          $\alpha$ & $\beta$ (GPa$^{-1}$)       & $\alpha$  & $\beta$ (GPa$^{-1}$)\\
\hline\hline
$C_{11}$ & 4.187    & 0.006         & 4.729     & $-$0.041\\ \hline
$C_{33}$ & 4.813    & $-$0.078        & 4.823     & $-$0.080\\ \hline
$C_{44}$ & $-$0.406   & 0.010         & 0.833     & $-$0.154\\ \hline
$C_{66}$ & $-$0.278   & $-$0.021        & $-$0.068    & $-$0.004\\ \hline
$C_{12}$ & 4.928    & $-$0.013        & 5.315     & $-$0.035\\ \hline
$C_{13}$ & 4.990    & $-$0.025        & 5.412     & $-$0.063\\
\hline\hline
\end{tabular}
\end{table}

The shear anisotropic factor for the (001) shear plane between [110] and [010] directions is defined in the case of tetragonal symmetry as follows \cite{He2011}:
\begin{equation}
A=\frac {2C_{66}}{C_{11}-C_{12}}\,.
\end{equation}

From table~\ref{tab:4}, the computed values of $A$ indicate that the elastic anisotropy for \{001\} shear planes between  $\langle 110\rangle$ and $\langle 010\rangle$ directions is somewhat higher in the case of CHTSe compared to CCTSe compound.

\begin{table}[!t]
\centering
\caption{Calculated elastic parameters: $A$, $B$, $G$, $E$, $\lambda$, $\nu$ and $B/G$ ratio for the ST-type of CCTSe and CHTSe adamantine compounds ($B$, $G$, $E$ and $\lambda$ in GPa, however, $A$, $\nu$ and $B/G$ are dimensionless).}
\label{tab:4}\vspace{2ex}
\begin{tabular}{|l|l|l|l|l|l|l|l|l|l|l|l|l|l|l|l|l|}
\hline\hline
&       $A$     & $B$     &   $G$     & $E$        & $\lambda$   & $\nu$   & $B/G$      \\
\hline\hline
CCTSe & 2.476   & 64.314  & 21.110    & 57.084     & 50.241      & 0.352      & 3.047           \\\hline
CHTSe & 2.546   & 63.174  & 22.142    & 59.477     & 48.412      & 0.343       & 2.853         \\
\hline\hline
\end{tabular}
\end{table}

Based on the calculated values of elastic constants ($C_{ij}$) for the single crystal, the estimation of the bulk modulus ($B$) and shear modulus ($G$) for polycrystalline substances is quite achievable through the Voigt-Reuss-Hill averaging scheme \cite{Hill1952}. Through both $B$ and $G$, a lot of mechanical parameters are worth to be investigated. Among them: Young's modulus ($E$), Lame's coefficient ($\lambda$) and Poisson's ratio ($\nu$). These parameters are given for polycrystalline substances by \cite{Grimvall1999}:
\begin{align}
E&=\frac{9BG}{3B+G}\,,\\
\lambda&=B-\frac{2}{3} G,\\
\nu&=\frac{3B-2G}{6B+2G}\,.
\end{align}

The obtained values are illustrated in table~\ref{tab:4}. The bulk modulus in the case of CCTSe is slightly higher than that of CHTSe, and the vice versa is true concerning the shear modulus. The estimated mechanical parameters are important in material sciences engineering, although in our study they are exploited to probe the thermal properties in the next section.

According to Pugh's empirical relationship \cite{Pugh1954}, if $B/G$ ratio is greater than 1.75, the material behaves as ductile, otherwise the material behaves as brittle. The $B/G$ ratios for the CCTSe and CHTSe compounds are 3.047 and 2.853, respectively, classifying both materials as ductile. Moreover, the Cd-based compound is more ductile than the Hg-based one, inter alia, these materials are appropriate for flexible samples applications.

\subsection{Thermal properties}

The Debye temperature is a primordial parameter in solid-state physics. Because of its relevance to the thermal response of crystals \cite{Anderson1963} it is used as a distinctive temperature between high and low temperature limits for solids \cite{Sun2012}. Using the average sound velocity ($v_{\text m}$), the Debye temperature ($\theta_{\text D}$) may be expressed as follows \cite{Anderson1963}:
\begin{eqnarray}
\theta_{\text D}=\frac{h}{k_{\text B}}\left(\frac{3n}{4\pi} \cdot \frac{N_{\text A} \rho}{M}\right)^{1/3}v_{\text m}\,,
\end{eqnarray}
here, $h$ is the Plank's constant, $k_{\text B}$ is the Boltzmann's constant, $N_{\text A}$ is the Avogadro's number,  $\rho$ is the density, $M$ is the molecular weight and $n$ is the number of atoms in the molecule.
For polycrystalline materials, ($v_{\text m}$) is related to the transverse $v_{\text t}$ and longitudinal $v_{\text l}$ sound velocities by \cite{Anderson1963}:
\begin{eqnarray}
v_{\text m}=\left[\frac{1}{3}\left(\frac{1}{v_{\text t}^3} + \frac{1}{v_{\text l}^3}\right)\right]^{-1/3},
\end{eqnarray}
where $v_{\text t}$ and $v_{\text l}$ are given by Navier's relations \cite{Anderson1963}:
\begin{eqnarray}
\left\{
\begin{array}{l}
\displaystyle v_{\text t}=\left(\frac{3B+4G}{3\rho}\right)^{1/2},\vspace{1ex}\\
\displaystyle v_{\text l}=\left(\frac{G}{\rho}\right)^{1/2}.
\end{array}
\right.
\end{eqnarray}

The calculated $\rho$, $v_{\text t}$, $v_{\text l}$, $v_{\text m}$ and $\theta_{\text D}$ are given in table~\ref{tab:5}. The obtained results predict that, $\theta_{\text D}$ of Cd-based compound is higher than that of Hg-based one. On the other hand, to the best of our knowledge, there are no data available in the literature on these properties for the considered materials in the herein study.
\begin{table}[!t]
\centering
\caption{Calculated density ($\rho$), transverse sound velocity ($v_{\text t}$), longitudinal sound velocity ($v_{\text l}$), average sound velocity ($v_{\text m}$), Debye temperature ($\theta_{\text D}$) and the thermal conductivity ($\kappa_{\text{min}}$) for both Cu$_2$\RN{2}SnSe$_4$ compounds considered.}
\label{tab:5} \vspace{2ex}
\begin{tabular}{|l|l|l|l|l|l|l|l|l|}
\hline\hline
& $\rho$ (g/cm$^3$) & $v_{\text t}$ (m/s) & $v_{\text l}$ (m/s) & $v_{\text m}$ (m/s) & $\theta_{\text D}$ (K) & \multicolumn{3}{c|}{$\kappa_{\text{min}}$ (W/mK)}\\
\cline{7-9}
&                   &             &             &             &                & \multicolumn{2}{c|}{This work}  & Experiment\\
\cline{7-8}
& &                  &               &                 &                   &   Clark & Cahill &\\
\hline\hline
CCTSe & 5.862            & 3971       & 1898       & 2563       & 265           & 0.452   & 0.661  & 0.29$-$1.01$^a$  \\\hline
CHTSe & 6.580            & 3753       & 1834       & 2467       & 254.5         & 0.434   & 0.624  & 0.5$^b$    \\
\hline\hline
\multicolumn{9}{l}{$^a$ references \cite{MLLiu2009, Ibanez2012}, $^b$ reference \cite{Li2013}.}
\end{tabular}
\end{table}

The thermal conductivity was evaluated using two theoretical methods: Clarke's \cite{Clarke2003} and Cahill's \cite{Cahill1992} model, which are given by the following equations:
\begin{equation}
\kappa_{\text{min}}^{\text{Clarke}}=0.87k_{\text B}\overbar{M}^{-2/3}E^{1/2}\rho^{1/6},
\end{equation}
here, $\overbar{M}=M/(nN_{\text A})$ is the average mass per atom, where $n$ is the number of atoms per formula.
\begin{equation}
\kappa_{\text{min}}^{\text{Cahill}}=\frac{k_{\text B}}{2.48}p^{2/3}\big(v_{\text l}+2v_{\text t}\big),
\end{equation}
here, $p$ is the density of the number of atoms per volume.

Table~\ref{tab:5} illustrates the calculated minimum thermal conductivities. The obtained values agree with the experimental ones. Thus, both exploited equations are authentic models for probing the thermal conductivity of these materials.

The lattice contribution to the heat capacity at a constant volume ($C_v$) is expressed as follows \cite{CASTEPOnlineHelp}:
\begin{equation}
C_v=k_{\text B} \int_{0}^{\infty} \frac{[\hbar\omega/(k_{\text B}T)]^2\re^{\hbar\omega/(k_{\text B}T)}}{[\re^{\hbar\omega/(k_{\text B}T)}-1]^2}g(\omega)\rd \omega,
\end{equation}
here, $\omega$ is the frequency, $g(\omega)$ is the normalized phonon density of states; $\int_{0}^{\infty}g(\omega)\rd \omega=1$ and $\hbar$ is the Planck's reduced constant.

\begin{figure}[!b]
\vspace{-4ex}%
\centerline{\includegraphics[height=0.28\textheight]{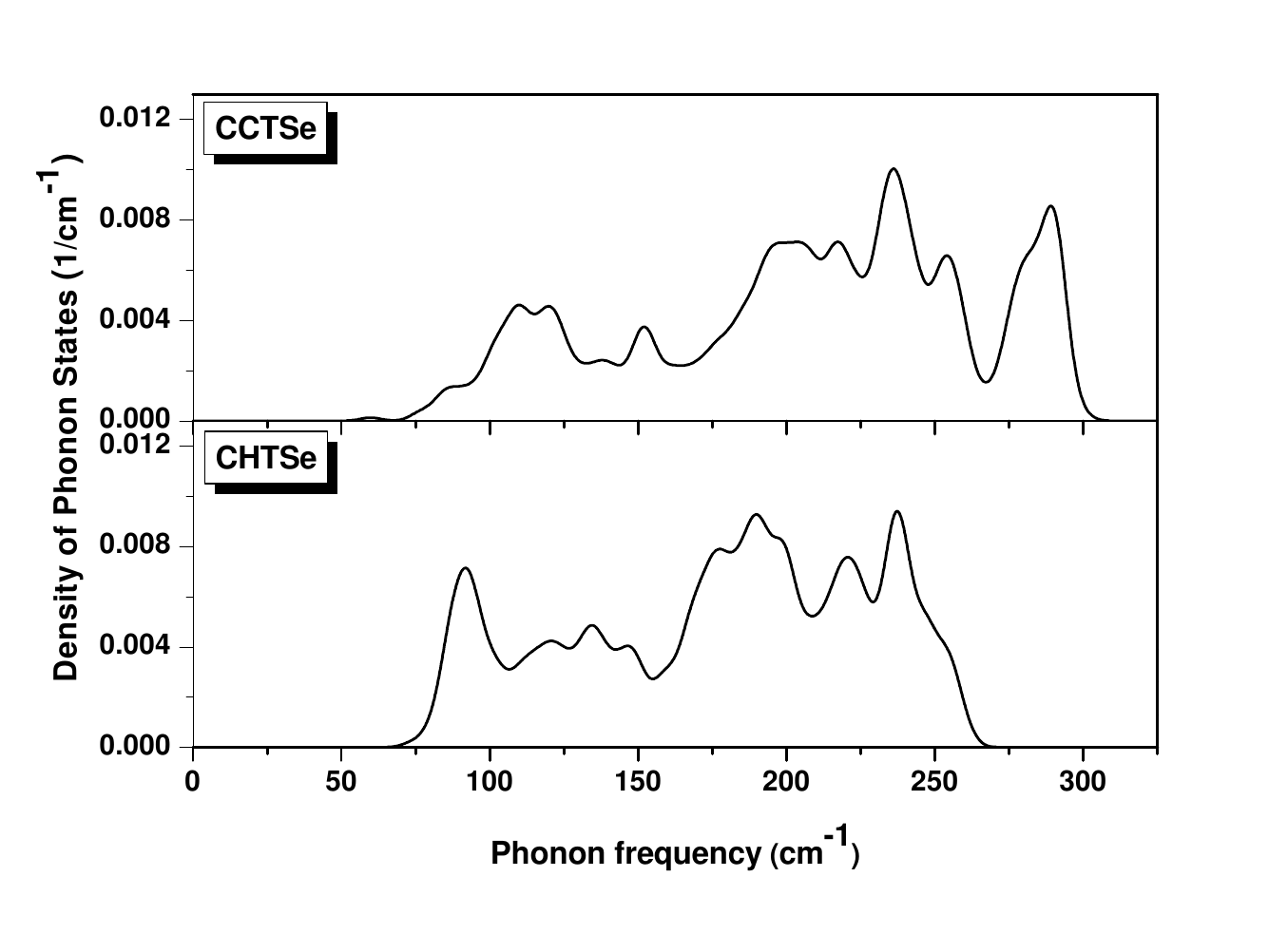}}
\vspace{-2ex}
\caption{Calculated density of phonon states for CCTSe and CHTSe adamantine materials.}
\label{fig-4}
\end{figure}

The quasi-harmonic approximation can lead to inaccurate results for temperatures too much exceeding the Debye temperature. In addition, CCTSe and CHTSe melt at 1053~K \cite{Matsushita2000} and 970~K \cite{Olekseyuk2002}, respectively. Therefore, to get significant curves, the temperature range 0--750~K is adopted.

\begin{figure}[!t]
\vspace{-2ex}%
\centerline{\includegraphics[width=0.7\textwidth]{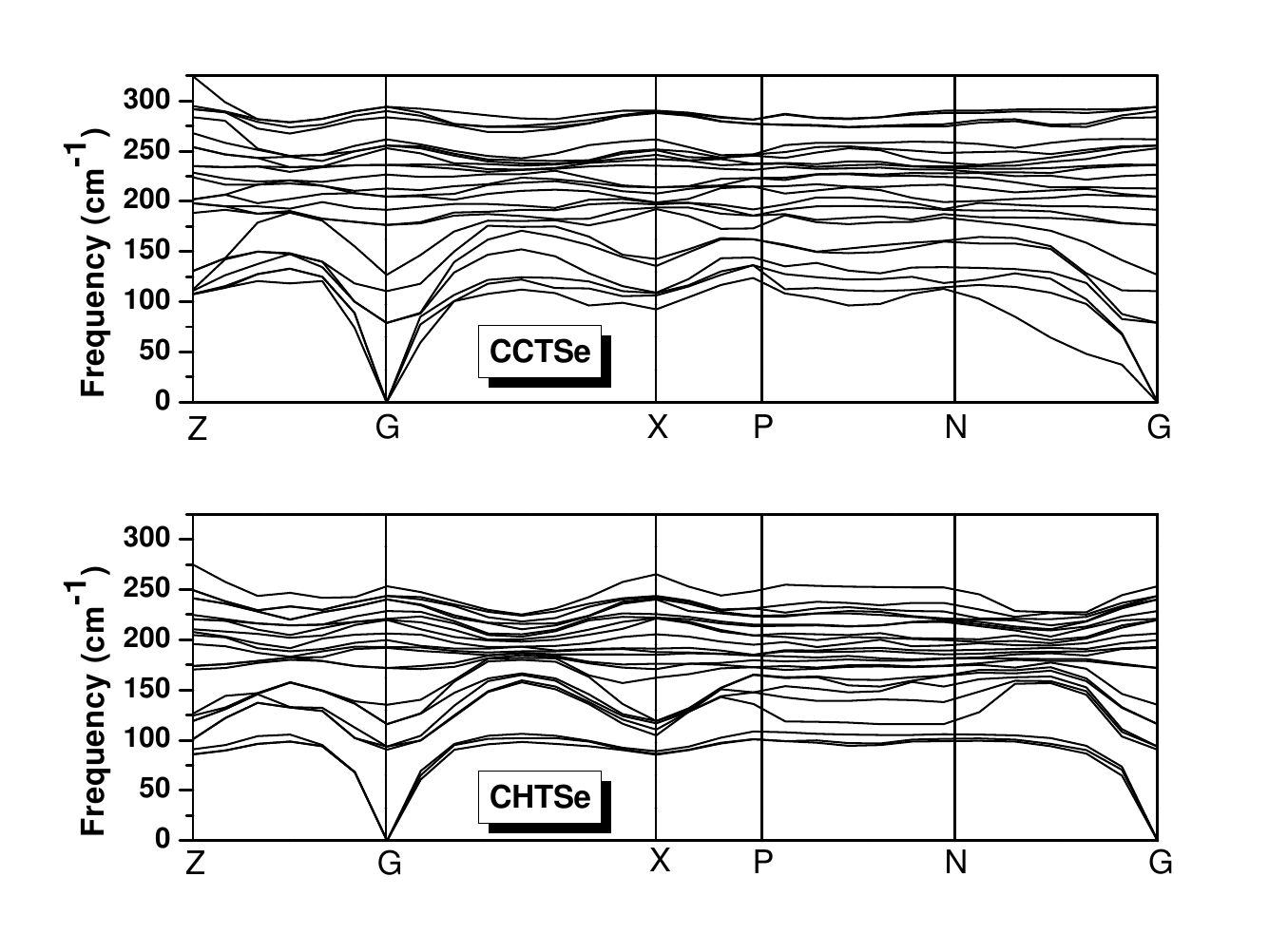}}
\vspace{-1ex}
\caption{Calculated phonon-dispersion curves along symmetry lines for CCTSe and CHTSe adamantine materials.}
\label{fig-5}
\end{figure}

The density of phonon states and the phonon dispersion spectra along the principal symmetry lines of the Brillouin zone (BZ) for both studied materials are displayed in figures~\ref{fig-4} and \ref{fig-5}, respectively. As figure~\ref{fig-5} shows, all phonon frequencies are positive, which means that both crystals are dynamically stable throughout the BZ.
\begin{figure}[!b]
\vspace{-4ex}%
\centerline{\includegraphics[width=0.7\textwidth]{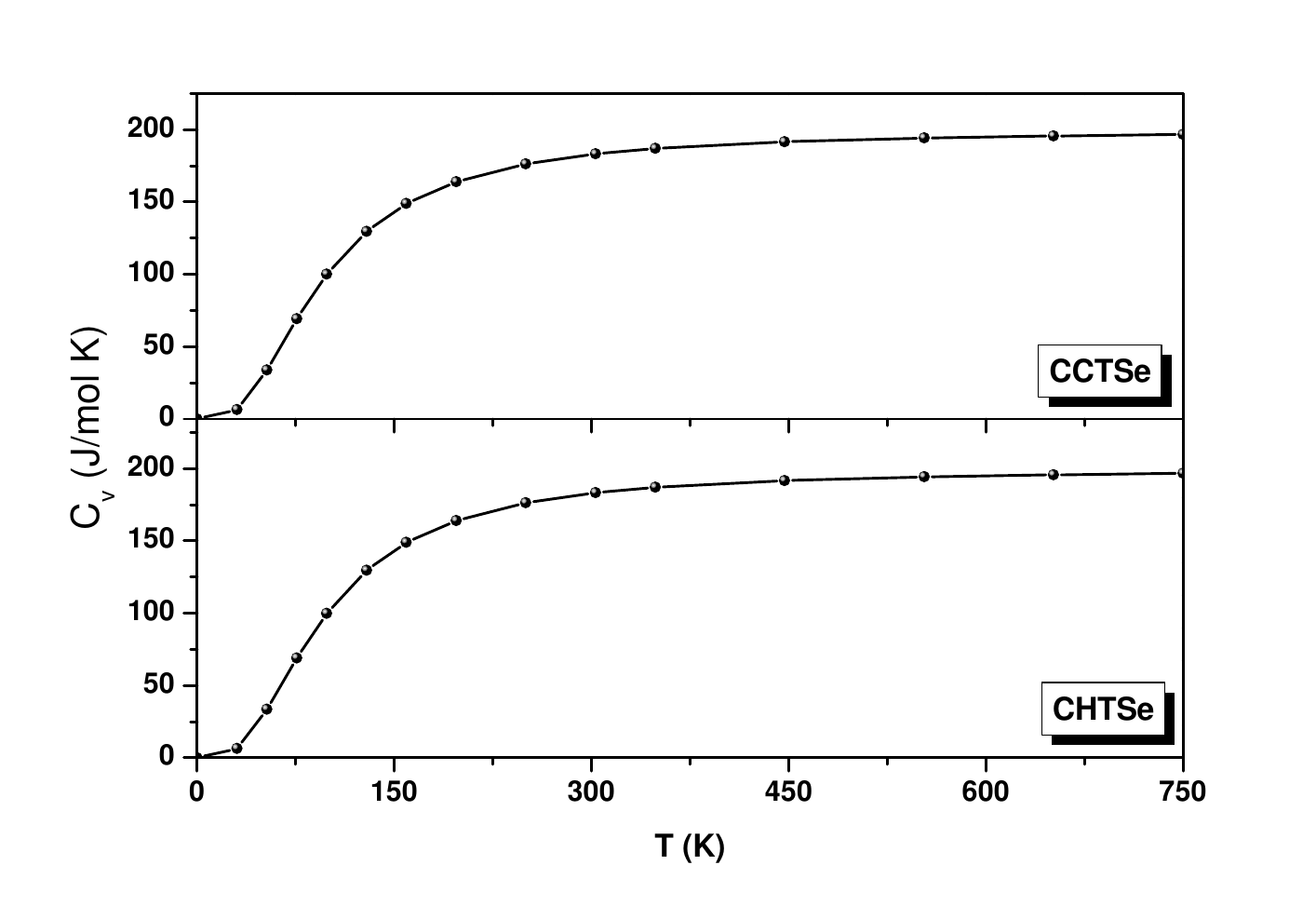}}
\vspace{-1ex}
\caption{Calculated heat capacity at constant volume, $C_v$, vs. temperature for CCTSe and CHTSe adamantine materials.}
\label{fig-6}
\end{figure}

As figure~\ref{fig-6} shows, an increasing temperature leads to a rapid increase of $C_v$ at lower temperature, then to a slow increase at high temperature and converges to a constant value. It is clear that when $T<300$~K, $C_v$ is proportional to $T^3$ \cite{Debye1912}. However, at high temperatures, it tends to the Petit and Dulong limit \cite{Dulong1819}, which is around ``197~J/mol\,K'' for both considered materials. Unfortunately, there are no available values of this parameter for the studied compounds. Hence, our calculated value could be taken as prediction.

 \section{Conclusion}
 \label{4}
In this paper, a computational insight on the structural, mechanical and thermal properties of CCTSe and CHTSe adamantine compounds has been achieved by first-principles calculation. The obtained optimized structures using GGA-WC approximation are in good agreement with the reported data in the literature. The probing of the elastic constants behavior under an external hydrostatic pressure, foresees that CCTSe and CHTSe are mechanically stable up to 10~GPa. The analysis of $B/G$ values shows that the stannite-type of both considered materials behaves as ductile, which suggests that these materials are appropriate for flexible samples. Using the calculated phonon density of states, the thermal behavior of the heat capacity is determined within the quasi-harmonic approximation. The heat capacity, $C_v$, exhibits similar variation vs. temperature and tends to ``197~J/mol\,K'' at high temperature for both studied crystals. Our fundamental contribution may enhance our knowledge about some mechanical and thermophysical aspects of these compounds. Therefore, the obtained results could be taken as a numerical aid for designing the CCTSe and CHTSe based devices.

\ukrainianpart

\title{Структурні, механічні і термічні властивості діамантових матеріалів Cu$_2$CdSnSe$_4$ і Cu$_2$HgSnSe$_4$
з першопринципних обчислень}
\author{С. Бенсалем\refaddr{label1},
М. Шегаар\refaddr{labe12,labe22},
A.~Бухемаду\refaddr{labe13}}
\addresses{
\addr{label1} Центр розвитку відновлювальної енергетики, 16340 Алжир, Алжир
\addr{labe12} Фізичне відділення, факультет природничих наук, Університет м. Сетіф, 19000 Сетіф, Алжир
\addr{labe22} Лабораторія оптоелектроніки і композитів, Університет м. Сетіф, 19000 Сетіф, Алжир
\addr{labe13} Лабораторія розробки нових матеріалів і їх характеристики, Університет м. Сетіф, 19000 Сетіф, Алжир}

\makeukrtitle

\begin{abstract}
Використовуючи першопринципні обчислення, що базуються на теорії функціоналу густини в рамках підходу псевдопотенціал-плоска хвиля,
ми вивчаємо структурні, механічні і термічні властивості діамантових матеріалів Cu$_2$CdSnSe$_4$ і Cu$_2$HgSnSe$_4$.
Обчислені параметри гратки добре узгоджуються з експериментальними і теоретичними результатами. Константи пружності обчислено для обох
сполук, використовуючи схему статичних скінченних деформацій.
Дія гідростатичного тиску на константи пружності передбачає, що   обидва матеріали є стійкими аж до  10~ГПа.
Полікристалічні механічні властивості, а саме,  коефіцієнт анізотропії  ($A$),  об'ємний модуль  ($B$), модуль зсуву ($G$),
модуль Юнга ($E$), коефіціент Ламе ($\lambda$) і відношення Пуасона ($\nu$) були отримані з обчислень пружних констант
монокристалу.
Аналіз відношення  $B/G$ показує, що обидві сполуки поводяться  як пластичні.
Базуючись на обчисленнях механічних параметрів,  досліджено температуру Дебая і термічну провідність. В рамках квазігармонічного
наближення досліджено температурну залежність питомої теплоємності гратки обох кристалів.
\keywords перші принципи, структурні параметри, механічні характеристики, термічні властивості, Cu$_2$CdSnSe$_4$, Cu$_2$HgSnSe$_4$
\end{abstract}

\end{document}